\documentstyle[prb,aps,epsf,twocolumn]{revtex}
\newcommand{\eq}{\begin{equation}}
\newcommand{\ee}{\end{equation}}
\newcommand{\s}{{\sigma}}

\newcommand{\vrr}{{\vec{r}}}

\newcommand{\vq}{{\vec{q}}}

\def\a{\alpha}
\def\b{\beta}
\def\e{\epsilon}

\def\dd{d^{\dagger}}

\def\half{{1\over2}}

\def\rhob{{\bar \rho}}

\def\ua{\uparrow}
\def\da{\downarrow}
\def\eqa{\begin{eqnarray}}
\def\eea{\end{eqnarray}}
\def\prl{{Phys. Rev. Lett.}}
\def\prb{{Phys. Rev. {\bf B}}}

\begin{document}
\draft
\flushbottom
\twocolumn[
\hsize\textwidth\columnwidth\hsize\csname @twocolumnfalse\endcsname
\title{Finite Temperature Magnetism  in  Fractional
Quantum Hall Systems: Composite Fermion Hartree-Fock and Beyond  }
\author{  Ganpathy Murthy}
\address{
Department of Physics and Astronomy, University of Kentucky, Lexington, KY 40506}
\date{\today}
\maketitle
\tightenlines
\widetext
\advance\leftskip by 57pt
\advance\rightskip by 57pt

\begin{abstract}
Using the Hamiltonian formulation of Composite Fermions developed
recently, the temperature dependence of the spin polarization is
computed for the translationally invariant fractional quantum Hall
states at $\nu=1/3$ and $\nu=2/5$ in two steps. In the first step, the
effect of particle-hole excitations on the spin polarization is
computed in a Composite Fermion Hartree-Fock approximation. The
computed magnetization for $\nu=1/3$ lies above the experimental
results for intermediate temperatures indicating the importance of
long wavelength spin fluctuations which are not correctly treated in
Hartree-Fock. In the second step, spin fluctuations beyond
Hartree-Fock are included for $\nu=1/3$ by mapping the problem on to
the coarse-grained continuum quantum ferromagnet. The parameters of
the effective continuum quantum ferromagnet description are extracted
from the preceding Hartree-Fock analysis. After the inclusion of spin
fluctuations in a large-$N$ approach, the results for the
finite-temperature spin polarization are in quite good agreement with
the experiments.

\end{abstract}
\vskip 1cm

]
\narrowtext
\tightenlines
\section{Introduction}
\label{intro}

The fractional quantum Hall (FQH) effect\cite{fqhe-ex} has introduced
us to new, highly correlated, incompressible states\cite{laugh} of
electrons in high magnetic fields.  A unified understanding of all
fractions $\nu=p/(2sp+1)$ was achieved by the Composite Fermion
picture of Jain\cite{jain-cf}, in which the electrons are dressed by $2s$
units of statistical flux to form Composite Fermions
(CFs). At a mean field level, the CFs  see a reduced field
$B^*=B/(2sp+1)$, in which they fill $p$ CF-Landau levels (CF-LLs), and
exhibit the integer quantum Hall effect. 

Due to the small $g$ factor of electrons in GaAs, spins may not be
fully polarized in FQH states\cite{hal-spin,singlet25}. Transitions
between singlet, partially polarized, and fully polarized states
(based on gap measurements) have been observed for a number of
fillings\cite{clark,initialex,buckthought,duetal}, which can be
understood in terms of CF's with a spin\cite{jain-cf,duetal,wu}. The
transitions happen when an unoccupied CF-LL of one spin crosses the
occupied CF-LL of the opposite spin. 

While these low temperature measurements are in satisfactory agreement
with the ground states predicted in the CF picture\cite{wu}, in order
to understand the temperature dependence of the polarization $P(T)$
one has to consider all excited states as well. Detailed measurements
of $P(T)$ for the $\nu=1/3$ state have recently appeared in the
literature\cite{barrett,melinte}. It is well-known that the $\nu=1/3$
state is spontaneously polarized at $T=0$, even when the Zeeman
coupling $E_Z=g\mu B_{tot}$ is zero. In this it is analogous to the
$\nu=1$ state\cite{shivaji-skyrmion}, which has been extensively
studied theoretically\cite{largen,many-body,haussmann} and
experimentally\cite{skyrmion-ex}. There are, however, significant
differences between the two cases at finite $T$. 

In a recent paper, MacDonald and Palacios\cite{macd-pala} identified a
key qualitative feature that makes $\nu=1/3$ very different from
$\nu=1$. In the $\nu=1$ case the particle-hole excitations are very
high in energy compared to $E_Z$, and are frozen out at all low
temperatures of interest. Consequently, the $T$ dependence of $P$
comes mainly from spin wave excitations and their interactions. This
is the reason why long-wavelength effective theories such as the
continuum quantum ferromagnet\cite{largen} approach are
successful. However, for $\nu=1/3$, particle-hole excitations are on
the same scale of energy as $E_Z$, and cannot be ignored at any
$T$. MacDonald and Palacios use a simplified model to illustrate this
feature\cite{macd-pala}, but the model is not sufficiently detailed to
enable a calculation of $P(T)$ for a realistic sample, and is
difficult to extend to non-Laughlin fractions.

The goal of this paper is to describe a general analytical method for
approximately computing $P(T)$ for an arbitrary principal fraction for
realistic samples. To this end, an approximate hamiltonian formalism
in which Composite Fermion variables explicitly appear will be
used\cite{us1}. This hamiltonian approach is based on the field
theoretic idea of attaching flux to electrons by using a Chern-Simons
field\cite{gcs,gmcs,zhk,read1,lopez,kalmeyer,hlr}. The CF-Hamiltonian
approach has many features suited to the computation of the physical
properties of fractional Hall systems. The CFs see the effective
field, and fill CF-Landau levels (CF-LLs). In the principal fractions
$\nu=p/(2p+1)$ with which we will be concerned, the effective field is
$B^*=B/(2p+1)$, and the CFs fill $p$ CF-LLs. The energies of the
CF-LLs are controlled entirely by interactions\cite{us1}, which is a
correct feature of the physics of electrons in the lowest Landau
level. Note that this is {\it not} a theory in which Composite
Fermions are free. On the contrary, the theory is fully interacting,
with both the kinetic energies and the residual interactions of the
CFs being controlled by the electron-electron interaction. Finally,
the nonperturbative charge and dipole moment of the excitations appear
explicitly in the theory\cite{us1}. The fact that all these nontrivial
features are built into the theory raises the expectation that very
simple approximate treatments of this Hamiltonian theory (such as
Hartree-Fock) will suffice to produce reasonable numbers for physical
quantities. This expectation is indeed borne out by explicit
calculations\cite{single-part,scaling,me-us}. The hamiltonian
formalism will be presented briefly in Section II.

The Hamiltonian approach is particularly suited to the computation of
finite-temperature properties in fractional Hall systems. In the
Composite Fermion Hartree-Fock (CFHF) approximation, one
self-consistently finds the single-particle energies and occupations
of the various CF-LLs at any finite temperature. Since the energies of
the states are controlled entirely by interactions and occupations,
{\it these energies will be temperature-dependent}. This is a familiar
feature in other interacting many-body systems, such as the BCS
superconductor, where the single-particle gap is a collective effect,
and depends on temperature. Thus the Hamiltonian theory gives us
valuable information on the evolution of the collective state as a
function of temperature, which is then used in the mapping to the
effective theory. Once the occupations of the single particle CF-LLs
have been determined, the polarization is computed simply as the
difference of the total $\ua$-spin and $\da$-spin occupations. This
yields the CFHF prediction for the temperature dependence of the spin
magnetization. This procedure and the results for $\nu=1/3$ and
$\nu=2/5$ are described in Section III. A brief description of this
work has appeared previously\cite{pvst}.

It turns out that for the spontaneously polarized $\nu=1/3$ state the
CFHF prediction is higher than the experimental values for all
temperatures, with the discrepancy being considerable for intermediate
temperatures, as shown in Figure 1. To put it in the proper context,
the agreement between CFHF and experiment is considerably better than
for $\nu=1$, where HF does a very poor job of predicting the spin
polarization\cite{many-body}. 
\begin{figure}
\narrowtext
\epsfxsize=2.4in\epsfysize=2.4in
\hskip 0.3in\epsfbox{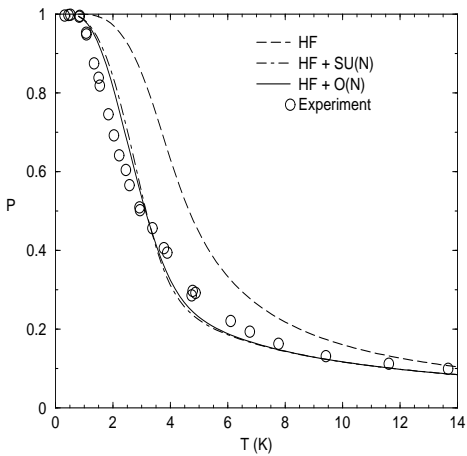}
\vskip 0.15in
\caption{Polarization versus $T$ for $\nu=1/3$. The circles are the 
data from the 10W sample of Khandelwal {\it et al}\cite{barrett}. The
dashed line is the prediction from the CFHF theory for a thickness
parameter of $\lambda=1.5l$. The solid and dot-dashed lines refer to
the theoretical prediction including spin fluctuations.
\label{fig1}}
\end{figure}
However, the discrepancy is nonetheless
there, and is presumably the result of spin fluctuations which are not
treated correctly in the  CFHF approximation.
One can imagine integrating out the fermions to obtain an effective
theory that has low energy spin degrees of freedom. One is then led to
map the low energy physics on to the continuum quantum ferromagnet
(CQFM)\cite{largen}. The CQFM has two free parameters, the
magnetization per unit volume $M_0$, and the spin stiffness
$\rho_s$. In the traditional CQFM theory these are
temperature-independent parameters. However, since the theory is
fermionic at the microscopic level, with the fermionic energy levels
and occupations being temperature dependent, one should expect $M_0$
and $\rho_s$ to acquire a $T$ dependence in the effective theory. It
turns out that these parameters can be easily extracted from the CFHF
treatment of Section III.  Armed with this information, we proceed to
include spin fluctuations in a large-N approach as described by Read
and Sachdev\cite{largen}. The results are in quite good agreement with
the experimental data over the whole range of temperature, as shown in
Figure 1 for the 10W sample of Khandelwal {\it et
al}\cite{barrett}. The same comparison is shown for the data of
Melinte {\it et al}\cite{melinte} in Figure 2. 
\begin{figure}
\narrowtext
\epsfxsize=2.4in\epsfysize=2.4in
\hskip 0.3in\epsfbox{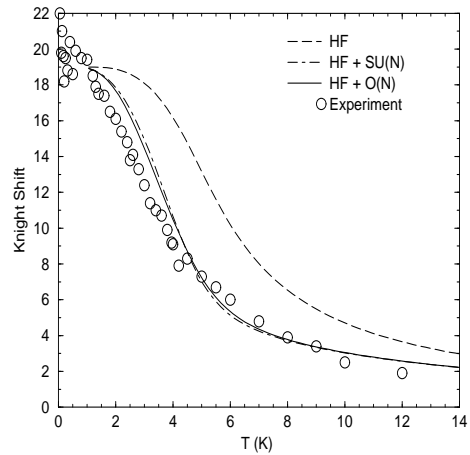}
\vskip 0.15in
\caption{The theoretical prediction in the CFHF approximation 
(dashed line) and including spin fluctuations (solid and dot-dashed
lines) compared to the Knight shift data from the M242 sample of
Melinte {\it et al}\cite{melinte}. A saturation value of $19kHz$ for
the Knight shift has been used, and the thickness paramter is assumed
to be $\lambda=1.5l$.
\label{fig2}}
\end{figure}

The details of the figures will be explained in Sections III and
IV. The mapping to the effective theory and the subsequent
calculations are described in Section IV.  We end with some
conclusions, caveats, and open questions.

\section{Hamiltonian Formalism}
\label{sec1}

Since detailed descriptions of the Hamiltonian theory of CFs have
appeared elsewhere\cite{us1}, we will restrict ourselves to a summary of the
essential features of this formalism. 

The Composite Fermion picture was originally used by
Jain\cite{jain-cf} to generate electronic wave functions with good
correlations. These wave functions have excellent overlap with the
exact wave functions for finite systems, and encode all the right
physics. In order to compute dynamical response functions, it is
desirable to have an operator or field theoretic formulation of
CFs. The fundamental property that CFs carry statistical flux can be
implemented by a Chern-Simons(CS) transformation, which
performs flux attachment via the CS gauge field to obtain a field
theoretic description with either bosons\cite{gcs,gmcs,zhk,read1} or
fermions\cite{lopez}. These theories have provided us with a link
between the microscopic formulation of the problem and experiment,
both for incompressible and compressible states\cite{kalmeyer,hlr}.

Recently R. Shankar and the present author developed a hamiltonian CS
theory for the fractional quantum Hall states\cite{us1}. Inspired
by the work of Bohm and Pines\cite{bohm-pines} on the 3D electron gas,
the Hilbert space was enlarged to introduce $n$ high-energy
magnetoplasmons degrees of freedom, ($n$ also being the number of
electrons) at the same time imposing an equal number of constraints on
physical states.  However, the fermions still had the bare mass
(recall that in the lowest Landau level, the electrons should lose all
memory of the bare mass, and acquire an effective mass controlled by
interactions), and the frequency of the magnetoplasmons was
incorrect. Hence a final canonical transformation was employed to
decouple the fermions from the oscillators {\em in the infrared
limit}.

	The final fermions are called {\em the} Composite Fermions for
the following reasons.  Firstly, the final fermions have no dispersion
in the absence of interactions and acquire an effective mass dependent
on interactions alone.  Next, the final canonical transformation
assigns to each fermion the magnetic moment $e/2m$ as mandated by very
general arguments\cite{ssh,simon-f1}. The central result of the
formalism is the formula for the electronic charge density, which
takes the following form, separable into high- and low-energy
pieces\cite{us1}, at small $q$:
\eqa
\rho_e(q)={q \over \sqrt{8\pi}} \sqrt{{2p\over2p+1}} (A(q)+
A^{\dagger}
(-q))&\nonumber\\
+{\sum_j e^{-iqx_j} \over 2p+1} -{il_{}^{2} }  (\sum_j (q \times
\Pi_j)e^{-iqx_j}&
)\label{rhobar}
\eea
where $A,A^{\dagger}$ refer to the annihilation and creation operators
of the magnetoplasmon oscillators, $l =1/\sqrt{eB}$ is the magnetic
length, and ${\vec\Pi}_j={\vec P}_j+e{\vec A}^*(r_j)$ is the velocity
operator of the CFs.  The oscillator piece saturates Kohn's
theorem\cite{kohn}. The rest, to be called $\rhob$, is obtained by
adding to the canonically transformed electronic charge density a
particular multiple of the constraint\cite{us1} (in the physical
subspace, one can add any multiple of the constraint without physical
consequences, but we wish to work in the full space). It has some very
useful properties in the full space:

\begin{itemize}
\item{} $\rhob$ satisfies the
magnetic translation algebra (MTA)\cite{GMP} to lowest leading
order. Since this is the algebra of the electron density in the lowest
Landau level (LLL), the LLL projection has been correctly carried out
in the infrared.
\item{} Note that $\rhob$ is a sum of a monopole with charge
$e^*=e/(2p+1)$, which is the charge associated with the CF, and a
dipole piece which alone survives at $\nu=1/2$ and has the value
proposed by Read\cite{read2}.  (A number of recent constructions have
emphasized this dipolar aspect\cite{dh,pasquier,read3,vonoppen}). 
\item{} We also find
that that as $\vq\to 0$ all transition matrix elements of $\rhob$ from
the HF ground state vanish at least as $q^2$.
\end{itemize}

The final property is an essential property of {\it physical } charge
density matrix elements from incompressible liquid ground states in
the LLL\cite{GMP}.  It is easy to see that if one intends to use the
Hartree-Fock approximation ignoring constraints, these properties of
$\rhob$ are essential. They make it plausible that $\rhob$ does not
suffer vertex corrections.

 The Hamiltonian of the low-energy sector  (dropping the magnetic
moment term)
is
\eq
H=\half \int {d^2 q\over(2\pi)^2} v(q) \rhob(-q) \rhob(q)\label{ham}.
\ee
where $v(q)$ is the electron-electron interaction.  Real samples 
  have a finite thickness $\Lambda$ of the same order as $l $, so that
  the Coulomb interaction is cutoff at large
  wavevectors\cite{thick1}. We will model this interaction by\cite{zds} 
\eq
v(q)={2\pi e^2\over q}\exp{-\lambda q}
\label{zdspot}\ee

Finally, the constraint will be approximately implemented by cutting
off the number of CF-LLs to maintain the correct number of {\it
electronic} states. For $\nu=p/(2p+1)$ this means keeping $2p+1$
CF-LLs. More details of the formalism can be found in refs.[19,29].

\section{The Composite Fermion Hartree-Fock Approximation}
\label{sec2}

The CFHF approximation has been applied to the above
Hamiltonian, and reasonable success has been obtained in computing
various physical quantities in the gapped
fractions\cite{single-part,scaling,me-us}, including a very recent
calculation of the temperature-dependent polarization $P(T)$ for the
compressible half-filled LL\cite{shankar12}. 

Before one employs the CFHF approximation, one needs to express the
Hamiltonian as an operator acting on a set of states. Since the CFs
see the effective field $B^*$, it is natural to represent the Hilbert
space as Slater determinants of single-particle CF-Landau level states
in the effective field. The wave functions of these states in the
Landau gauge are
 
\eq
\phi_{n,X}(\vrr) ={1\over\sqrt{L}} e^{iXy/(l^*)^2} \phi_n((x-X)/l^*)
\ee
Here $l^*=l\sqrt{2p+1}$ is the effective magnetic length, $L$ is the
linear dimension of the system, $X=2\pi (l^*)^2j/L$ is the degneracy
index, and $j=0, 1, . . . , L^2/(2\pi(l^*)^2)$, and $\phi_n$ is the
$n^{th}$ normalized harmonic oscillator eigenfunction. The index $n$
is the CF-LL index.

The density is a one-body operator, and can be expressed in this basis
(after the spin labels have been included) as
\eq
\hat{\rhob}(\vq)=\sum_{\s,X\{n_i\}} e^{-iq_xX} \rho_{n_1n_2}(\vq)
\dd_{\s,n_1,X-{q_y(l^*)^2\over2}}d_{\s,n_2,X+{q_y(l^*)^2\over2}}
\ee
where $\s=\ua, \da$ is the spin index, and $d_{\s,n,X}$ destroys a CF
in the single-particle state labelled by ${\s,n,X}$, and
$\rho_{n_1n_2}(\vq)$ is a matrix element given by 

\eqa
&\rho_{n_1n_2}(\vq)={(-1)^{n_<+n_2}\over
2p+1}\sqrt{n_<!\over
n_>!} e^{i(\theta_q-{\pi\over2})(n_1-n_2)} \bigl({ql^*\over\sqrt{2}}\bigr)^{n_>-n_<}\nonumber \\
&\times
e^{-y/2}(n_>L_{n_<-1}^{n_>-n_<}+2L_{n_<}^{n_>-n_<}-(n_<+1)L_{n_<+1}^{n_>-n_<})
\label{rhomat}
\eea
where $n_<$ ($n_>$) is the lesser (greater) of $n_1, n_2$, $\theta_q$
is the angle of the vector $\vq$ measured from the $x$-axis,
$y=(ql^*)^2/2$ is the argument of the Laguerre polynomials $L_n^k$.

The Hamiltonian is now a four-fermi operator. It can easily be shown
that the single-particle states defined above form a good HF
basis\cite{me-us}. In the CFHF approximation, one reduces the
four-fermi Hamiltonian to a two-fermi operator by taking averages
according to the rules
\eqa
&<GS|\dd_{\nu} d_{\nu'}|GS>=\delta_{\nu\nu'}N_F(\nu)\\
&<GS|d_{\nu} \dd_{\nu'}|GS>=\delta_{\nu\nu'} (1-N_F(\nu))
\eea
where $\nu$ is a shorthand for all the state labels. This results in
the HF Hamiltonian
\eq
H_{HF}=\sum_{\s,n,X} \e(\s,n) \dd_{\s,n,X}d_{\s,n,X}
\ee
where the HF single-particle energy is 
\eqa
&\e(\s,n)=\nonumber\\
&\pm {E_Z\over2}+\half\int{d^2q\over(2\pi)^2} v(q)\sum_{m}(1-2N_F(\s,m))|\rho_{nm}(q)|^2
\label{eformula}\eea
in  which the Zeeman energy  has been  added 
($E_Z=g^*\mu_B B$).

Finding the energies at $T=0$ is quite simple, since the occupations
of the CF-LLs can only be 0 or 1. For example, for the $\nu=1/3$
state, $N_F(\ua,0)=1$ and all other occupations are zero, and for the
$\nu=2/5$ singlet state $N_F(\ua,0)=N_F(\da,0)=1$, and all other
occupations are zero. For nonzero $T$ one has to carry out a
self-consistent procedure. First one chooses trial values for the
energies (say the $T=0$ values) and a trial value for chemical
potential $\mu$ (say halfway between the lowest unoccupied and highest
occupied CF-LL). Then one assigns the occupations of the
single-particle levels according to Fermi occupation function

\eq
N_F(\s,n)={1\over1+\exp{(\e(\s,n)-\mu)/T}}
\ee
with the trial value of $\mu$.  Then one recomputes the HF energies
using equation (\ref{eformula}). Note that the structure of degenerate
CF-LLs remains intact, and only the energies and occupations
change. Since the filling has to remain fixed, the chemical potential
will change when the energies change. One then recomputes the chemical
potential as the root of the equation
\eq
\sum_{n}(N_F(\ua,n)+N_F(\da,n))=p
\ee
for the principal fraction $p/(2p+1)$ and iterates the whole process
until self-consistency is acheived.  Finally, the spin polarization is
given by

\eq
P=\sum_{n}(N_F(\ua,n)-N_F(\da,n))/p
\ee

From the above procedure it is clear that only single-particle
excitations have been taken into account in obtaining $P(T)$ so far,
and spin fluctuations have been explicitly ignored. If it so happens
that for the system under consideration the effects of spin
fluctuations are small, then this approximation should be accurate,
otherwise not.

Let us proceed to compare the CFHF results to experiments. We first
consider the 10W sample of Khandelwal {\it et al}\cite{barrett}. The
sample parameters are $B_{\perp}=9.61 T$, and $B_{tot}=12T$. This
implies that the Coulomb energy scale is $E_C=e^2/\varepsilon
l\approx160 K$ and the Zeeman energy is $E_Z=0.0175 E_C$. We will use
a value of $\lambda=1.5l$ for the thickness parameter for illustrative
purposes throughout this paper. This value ought to be in the physical
regime\cite{morf} for most samples, and also approximately
agrees with that extracted from an analysis of the compressible
states\cite{shankar12}. {\it It should be emphasized that the CFHF
analysis, and the mapping to the effective spin theory that follows,
can be performed for any potential $v(q)$}.

Figure 1 (in the Introduction) shows the HF prediction from our theory
for $\lambda=1.5l$ compared to the experimental data. The agreement is
good at very low ($T\le 1K$) and very high ($T\ge 8K$)
temperatures. However, at intermediate $T$, there is a big discrepancy
between the CFHF prediction and the data. This indicates that effects
not treated correctly in HF, notably long wavelength thermal spin
fluctuations, are important in this intermediate regime of
$T$. Nonetheless, to put the result in context, one should note that
the CFHF prediction agrees much better with the data than the
corresponding HF prediction for $\nu=1$ (see, for example, ref[15] for
a comparison of the different predictions for $\nu=1$).  This is
because at $\nu=1$ particle-hole excitations are completely
unimportant at all temperatures of interest, while thermal spin
fluctuations dominate. Since long wavelength spin fluctuations are
treated very poorly in HF the agreement is bad. However, at
$\nu=1/3$, particle-hole excitations play a major role in reducing
$P(T)$ for $T>5K$. In the next section we will see how to incorporate
spin fluctuations into our calculation, resulting in much better
agreement with the data.

\begin{figure}
\narrowtext
\epsfxsize=2.4in\epsfysize=2.4in
\hskip 0.3in\epsfbox{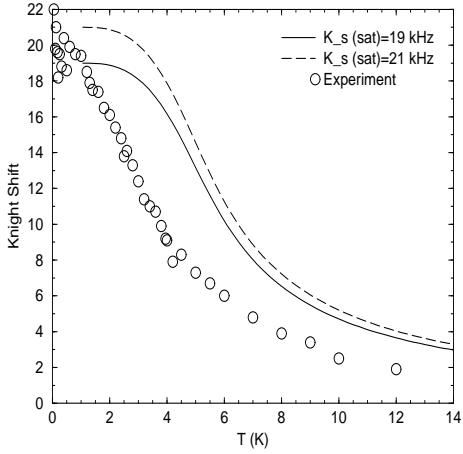}
\vskip 0.15in
\caption{Knight shift versus $T$ for $\nu=1/3$. The circles are the 
data from the M242 sample of Melinte {\it et al}\cite{melinte}. The
 lines are the predictions from the CFHF theory for $\lambda=1.5l$,
 assuming two different values for the Knight shift that corresponds
 to saturated polarization.
\label{fig3}}
\end{figure}

Figure 3 shows the same type of comparison for the Knight shift data
of Melinte {\it et al}\cite{melinte}for their M242 sample.  Here the
sample parameters are $B_{tot}=B_{\perp}=17T$. This implies that
$E_C\approx210K$, $E_Z=0.0186 E_C\approx4K$.  Once again a value of
$\lambda=1.5l$ has been used in the CFHF calculation. In order to
translate the Knight shifts into polarization numbers or vice versa,
one needs to determine the saturation value of the Knight shift, which
presumably corresponds to a polarization of $P=1$. There is a lot of
scatter in the data at low $T$, due to the very long times needed to
measure the Knight shift, and the error bars are also large at low
$T$\cite{melinte}. There is thus some uncertainty in determining the
Knight shift corresponding to $P=1$. In any reasonable theory one
expects to find that $P=1$ for $T\ll E_Z$, and expects to see this
saturated value of $P$ up to about $T=0.5E_Z$ or so.

Based on these considerations two values of the saturation Knight
shift $K_{s,P=1}$, $21kHz$ and $19kHz$ have been used to fit the data
here, both of which lie within the error bars of the low $T$
data\cite{melinte}. One possibility that can explain this spread is
that spin-reversed quasiparticles are present in the ground state due
to disorder, which can bring down the ``saturated'' value of the
Knight shift\cite{jain-pvt}. The $21kHz$ value was used by Melinte
{\it et al} in a phenomenological $\tanh(\Delta/4k_BT)$ fit to obtain
$\Delta=1.7E_Z$. However, one must note that the fit for
$K_{s,P=1}=19kHz$ seems slightly better, since then the experimental
saturation region is about $0.5E_Z$. The agreement between theory and
experiment for this value of $K_{s,P=1}$ are also better than for
$K_{s,P=1}=21kHz$. The $19kHz$ value will be used in the mapping to
the effective theory in the next section. Overall the agreement is
slightly worse than for the Khandelwal {\it et al} data\cite{barrett},
but leads to the same conclusion: It is quite important to treat
thermal spin fluctuations correctly at intermediate temperatures for
$\nu=1/3$.

Before we proceed to approximately incorporate spin fluctuations into
the theoretical prediction, let us address the following interesting
question: Why is HF so good in this case relative to the case of
$\nu=1$? To answer this question let us turn to the spin wave
dispersions. The spin wave is a collective spin-flip excitation, and
at wave vector $\vq$ corresponds a plane wave state in which a
majority spin quasihole and a minority spin quasiparticle are at a
separation of $q(l^*)^2$.
\begin{figure}
\narrowtext
\epsfxsize=2.4in\epsfysize=2.4in
\hskip 0.3in\epsfbox{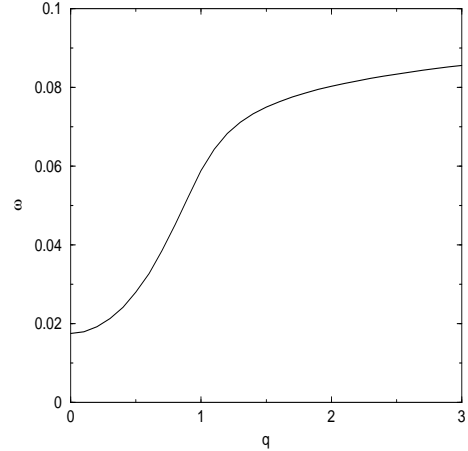}
\vskip 0.15in
\caption{Spin wave energy dispersions in units of the Coulomb energy $E_C$  
for $E_Z=0.0175$ and $\lambda=1.5l$, with $q$ plotted in units of
$l^{-1}$. As can be seen, the scale of the spin-reversed gap is the
same as $E_Z$.
\label{fig4}}
\end{figure}
The $q\to 0$ limit is required to be $E_Z$ by Larmor's theorem, while
in the $q\to\infty$ limit the particle and hole become infinitely
separated, so that the energy of the excitation is the spin-reversed
particle-hole gap $\Delta_{SR}$.  The dispersion of these excitations
can be computed for $\nu=1/3$ in the manner described in ref.[29], and
is shown in Figure 4 for $\lambda=1.5l$ for $E_Z=0.0175 E_C$ and
$T=0$. Figure 4 explicitly illustrates the feature\cite{macd-pala}
that the spin-flip particle-hole excitations are at the same energy
scale as $E_Z$.

How does this evolve as temperature increases?  Recall that all the
energy splittings in the CF-Hamiltonian formalism come from
interactions, and as the occupations of the states change with $T$ so
do their HF energies. As $T$ increases the occupations of the minority
spin levels increase while that of the lowest majority spin level
decreases. This means that the exchange splitting between the minority
and majority spin levels {\it decreases}, as can be seen from equation
(\ref{eformula}). It is clear that as $T$ becomes very large the
occupations of all the levels should tend to become the same, and
therefore $\Delta_{SR}$ should tend towards $E_Z$. Since the spin wave
dispersion has to be $E_Z$ for very small $q$, and $\Delta_{SR}$ for
very large $q$, this implies that the spin wave dispersion becomes
increasingly flat at $T$ increases. We will estimate the spin
stiffness in the next section and corroborate this conclusion.  Figure
5 shows this behavior of $\Delta_{SR}$ explicitly for the same
parameters as in Figure 4.

\begin{figure}
\narrowtext
\epsfxsize=2.4in\epsfysize=2.4in
\hskip 0.3in\epsfbox{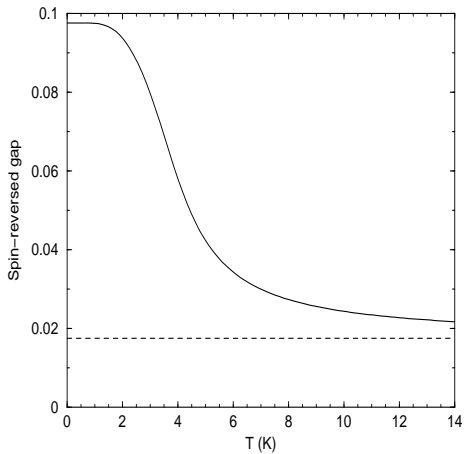}
\vskip 0.15in
\caption{Spin-reversed gap $\Delta_{SR}$ in units of $E_C$ as a function of $T$. 
The dashed line is $E_Z$.  Beyond about $6K$ $\Delta_{SR}$ is
essentially the same as $E_Z$.
\label{fig5}}
\end{figure}

Now consider a {\it noninteracting} theory of CFs. This would have a
completely flat dispersion for the collective mode, that is,
$\omega(q)=E_Z$.  Therefore, as the temperature increases, the theory
appears to be more weakly interacting, and the CFHF approximation
becomes more accurate. This trend can be expected to continue until a
temperature scale when CFs cease to exist. There are no obvious signs
of such a scale in the data.

The CF-Hamiltonian theory and the CFHF approximation are very general,
and can be applied to any fractional Hall state (for an application to
compressible states see ref[42]). To ilustrate this Figure 6 shows the
$P(T)$ curves for $\nu=2/5$ for $\lambda=1.5l$ for a range of Zeeman
couplings. Note the nonmonotonicity of the curves that start from the
singlet ground state at $T=0$. There is a transition to the fully
polarized state around $E_Z=0.01E_C$.
\begin{figure}
\narrowtext
\epsfxsize=2.4in\epsfysize=2.4in
\hskip 0.3in\epsfbox{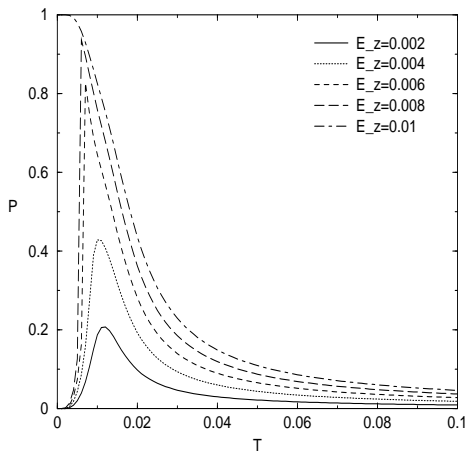}
\vskip 0.15in
\caption{Polarization versus $T$ for $\nu=2/5$ and $\lambda=1.5l$, 
for a range of values of $E_Z$. $E_Z$ and $T$ are both plotted in
units of $E_C$.
\label{fig6}}
\end{figure}
Note
also that only results for translationally invariant HF states are
presented, which ignores possible partially polarized {\it
inhomogeneous} states that have recently been proposed\cite{ppdw} to
explain intriguing observations by Kukushkin {\it et al}\cite{kukush}
of a state with half the maximal polarization for $\nu=2/5$, which is
not allowed as a translationally invariant CF state at $T=0$.

Let us compare our results to the only other method that can
compute $P(T)$ for arbitrary fractions, which is exact diagonalization
(keeping all the excited states) and subsequent calculation of
thermodynamic quantities\cite{exact}. Due to computational
limitations, this method is restricted to fairly small systems. For
example, the largest system studied by Chakraborty and
Pietilainen\cite{exact} for $\nu=1/3$ has 5 electrons, and for
$\nu=2/5$ has 4 electrons.  For $\nu=1/3$ the exact diagonalization
result lies above our predictions (and the experiment) for
$T>4K$. This discrepancy might be the result of finite thickness
and/or finite size corrections. However, at low $T$ the exact
diagonalization result\cite{exact} follows the data more closely than
our HF approximation (in all the above comparisons the $g=0.5$ line in
Figure 2 of ref[48] has been used and compared to the 10W sample of
Khandelwal {\it et al}\cite{barrett}. This sample has the closest
parameters to those used in ref[48]). For $\nu=2/5$, the CFHF  results
reproduce the nonmonotonicity of $P(T)$ for those values of $E_Z$
where the singlet state is the ground state, and the peaks in $P(T)$
occur at roughly the same $T$ in the CFHF and the exact
diagonalization results\cite{exact}. However, the same overall pattern
holds for $\nu=2/5$, namely, the results of Chakraborty and
Pietilainen\cite{exact} are below the CFHF prediction for low $T$, but
are higher for $T>0.02E_C$, where they once again see a $1/T$ tail
with a large coefficient. It would be interesting to explore the
finite size systematics to see if the large $T$ tail is suppressed for
larger sizes.

\section{Spin Fluctuations: Mapping to the Continuum Quantum Ferromagnet}
\label{sec3}

As can be seen in Figures 1 and 2, spin fluctuations are quite
important for $\nu=1/3$ at intermediate temperatures. for $\nu=1$, the
coarse-grained effective theory of the continuum quantum ferromagnet
(CQFM) coupled with the large-$N$ approximation, first applied to this
problem by Read and Sachdev\cite{largen}, has been quite successful in
explaining the temperature dependence of the polarization. In this
section a method is presented to map the $\nu=1/3$ problem to the
CQFM.

In the CQFM description one starts with the action\cite{largen}
\eqa
S=&\int d^dx\int\limits_{0}^{1/T} d\tau (iM_0
\vec{A}(\vec{n})\cdot\nabla_{\tau}\vec{n} + {\rho_s\over2}(\nabla_{x}\vec{n})^2 \nonumber\\
&- M_0 \vec{H}\cdot\vec{n}+\cdots)
\eea
where $\vec{n}$ is a local vector of unit length pointing in the
direction of the magnetization, $\vec{A}(\vec{n})$ is the field that
implements the Berry's phase needed to obtain the correct quantum
commutation relations between the spin components, and
$\vec{H}=g^*\mu_B \vec{B}$ is the Zeeman field ($|\vec{H}|=E_Z$). The
two crucial parameters which enter the action are $M_0$, the
magnetization density, and $\rho_s$, the spin stiffness. There are
other omitted terms in the action, which produce at most a logarithmic
correction to the magnetization in two dimensions\cite{largen}. There
are various ways one can proceed at this point\cite{kopietz}, but the most
convenient one for our purposes involves mapping the spin problem to a
problem with Schwinger bosons\cite{assa-book}, and making the large-$N$
approximation\cite{assa-book}. There are two common ways of mapping the
spin problem to Schwinger bosons: The $SU(N)$ approach and the $O(N)$
approach\cite{largen}, which give slightly different answers.

In the following, we will restrict the theoretical results to the
leading large-$N$ approximation, where the magnetization is given as
$M(T)=M_0\Phi_M(r,h)$, where $r=\rho_s/T$ and $h=E_Z/T$ are scaling
variables, and $\Phi_M$ is a scaling function. In this approximation,
the results of Read and Sachdev\cite{largen} for the magnetization in
the $SU(N)$ approach are
\eq
\Phi_M(r,h)={\log(q_1-e^{-h/2})-\log(q_1-e^{h/2})\over8\pi r}
\label{scales}
\ee
 and $q_1>1$ is
a root of the equation
\eq
(q_1-e^{-h/2})(q_1-e^{h/2})=q_1^2e^{-8\pi r}
\ee

The corresponding results in the leading $O(N)$
approximation\cite{largen} are
\eq
\Phi_M(r,h)={\log(q_2-e^{-h})-\log(q_2-e^{h})\over4\pi r}
\label{scaleo}
\ee
where $q_2>1$ is the solution of 
\eq
(q_2-e^{-h})(q_2-1)(q_2-e^{h})=q_2^3e^{-4\pi r}
\ee

The principal assumption underlying the CQFM description is that
long-wavelength ferromagnetic spin fluctuations are the only low
energy modes that affect the spin polarization in the temperature
range of interest. In the regime where this assumption holds for the
$\nu =1/3$ state, it should be possible to map the CF-Hamiltonian
theory to the CQFM. Conceptually, one can think of ``integrating out''
the fermions and leaving behind an effective theory of the spin
fluctuations. Operationally, one needs to find the values of $M_0$ and
$\rho_s$ corresponding to the $\nu=1/3$ state. An additional
complication arises here: Since the underlying fermionic theory
responds to temperature by self-consistently modifying occupations and
energies, one should expect to obtain temperature dependent values
$M_0(T)$ and $\rho_s(T)$. We will extract these values from the CFHF
Hamiltonian results.

First consider $M_0(T)$. We already have a value for the CFHF
magnetization $M_{HF}(T)$. Let us first write down the correct relation between
$M_0$ and $M_{HF}$ and then justify it.
\eq
M_{HF}(T)=M_0(T)\Phi_M(r=0,h=\Delta_{SR}/T)
\label{m0}\ee

In the CFHF theory the particles and holes are treated as independent,
or noninteracting, with a gap equal to $\Delta_{SR}$ (to be more
precise, this is the lowest energy spin-flip excitation). To put it
differently, the energy gap to create this spin-flip excitation is
always $\Delta_{SR}$, no matter what the distance between the particle
and the corresponding hole. This is as though the collective mode
dispersion were completely flat, $\omega(q)=\Delta_{SR}$. The CQFM
description that corresponds most closely to the CFHF is the one that
has {\it the same spin-flip excitation spectrum}, namely one with no
spin stiffness, that is, $r=0$, and and an effective Zeeman field
$E_Z^{eff}=\Delta_{SR}$. The CQFM prediction for the magnetization of
such a theory is the right hand side of equation (\ref{m0}), which has
to be equated to the CFHF prediction, hence the above formula, equation
(\ref{m0}).

Knowing that, for $SU(N)$,
\eq
\Phi_M(0,h)=\tanh(h/2)
\ee
while, for $O(N)$,
\eq
\Phi_M(0,h)={\sinh(h)\over1/2+\cosh(h)}
\ee
one extracts $M_0(T)$ from equation (\ref{m0}), using the value of
$M_{HF}(T)$ computed in Section III\cite{itinerant}.

Next consider the spin stiffness $\rho_s(T)$. At a given temperature
$T$ the self-consistent occupations $N_{F,GS}(\s,n)$ and energies
$\e(\s,n)$ in the ground state are computed using the procedure
described in Section II. Now one creates a twisted spin state by
defining
\eqa
d_{\ua,n,X}=&\cos(qX/2) d_{\a,n,X} + \sin(qX/2) d_{\b,n,X}\nonumber\\
d_{\ua,n,X}=&\cos(qX/2) d_{\a,n,X} + \sin(qX/2) d_{\b,n,X}
\eea
where $\a,\b$ define {\it local} directions of up and down.  In the
twisted spin state the occupations of the local up and down spins
remain the same as in the ground state, that is
\eqa
<\dd_{\a,n,X}d_{\a,n',X'}>=&\delta_{nn'}\delta_{XX'} N_{F,GS}(\ua,n)\nonumber\\
<\dd_{\b,n,X}d_{\b,n',X'}>=&\delta_{nn'}\delta_{XX'} N_{F,GS}(\da,n)\nonumber\\
<\dd_{\a,n,X}d_{\b,n',X'}>=&0
\eea

This leads to the following expectation values for the actual (global)
$\ua$ and $\da$ spin directions.

\eqa
<\dd_{\ua,n,X}d_{\ua,n',X'}>=&\delta_{nn'}\delta_{XX'}\times\nonumber\\
(\cos^2(qX/2)N_{F,GS}(\ua,n)&+\sin^2(qX/2)N_{F,GS}(\da,n))\nonumber\\
<\dd_{\da,n,X}d_{\da,n',X'}>=&\delta_{nn'}\delta_{XX'}\times\nonumber\\
(\sin^2(qX/2)N_{F,GS}(\ua,n)&+\cos^2(qX/2)N_{F,GS}(\da,n))\nonumber\\
<\dd_{\ua,n,X}d_{\da,n',X'}>=&\delta_{nn'}\delta_{XX'}\times\nonumber\\
\sin(qX/2)\cos(qX/2)&(N_{F,GS}(\ua,n)-N_{F,GS}(\da,n))\nonumber\\
<\dd_{\da,n,X}d_{\ua,n',X'}>=&<\dd_{\ua,n,X}d_{\da,n',X'}>
\label{twist}\eea
and corresponds to a state where the unit vector $\vec{n}$ pointing in
the direction of the local magnetization has components
$\vec{n}=(\sin(qX),0,\cos(qX))$.  Using the expectation values from
equation (\ref{twist}), one can compute the HF energy of the twisted
ground state, and thence the excess energy to order $q^2$.  Comparing
to the energy cost of a twist in the CQFM, which is $(\rho_s/2)
L^2q^2$, one finds the spin stiffness
\eqa
\rho_s=&{1\over16\pi}\int{d^2s\over(2\pi)^2} v(s) 
\sum\limits_{n_1,n_2} |\rho_{n_1n_2}(s)|^2\times\nonumber\\
& (N_F(\ua,n_1)-N_F(\da,n_1))(N_F(\ua,n_2)-N_F(\da,n_2))
\label{rhos}\eea
where $L^2$ is the area of the system, and $\rho_{n_1n_2}$ is the
density matrix element of equation (\ref{rhomat}). One caveat should
be mentioned here: The above should be regarded as an estimate for the
twist rather than a rigorous calculation (even in HF), since ideally
one should compute the {\it free energy} cost of a twist, rather than
just the {\it internal energy} cost, as we have done. The free energy
cost would be computed by carrying out a self-consistent HF at finite
$T$ in the presence of a twist. However, a fully relaxed HF
calculation of an inhomogeneous state at $T\ne0$ is computationally
prohibitive. In the following we use the value given by equation
(\ref{rhos}).

\begin{figure}
\narrowtext
\epsfxsize=2.4in\epsfysize=2.4in
\hskip 0.3in\epsfbox{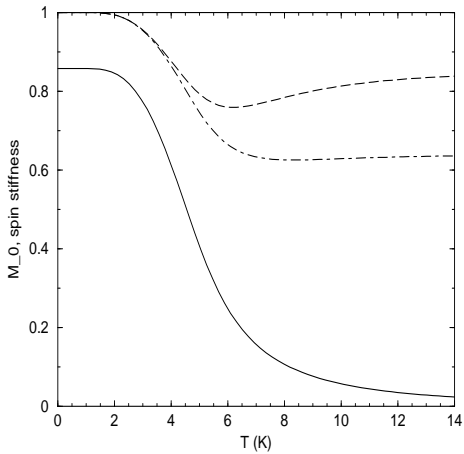}
\vskip 0.15in
\caption{The parameters $M_0(T)$ (normalized to 1 at $T=0$) for 
$SU(N)$ (dashed line) and $O(N)$ (dot-dashed line), and $\rho_s(T)$
($\times 400$ in units of $E_C$) extracted from the CFHF for
$E_Z=0.0186 E_C$ and $\lambda=1.5l$. While $M_0$ shows no dramatic
behavior, $\rho_s$ decreases precipitously beyond $3K$.
\label{fig7}}
\end{figure}

Figure 7 shows the results for $M_0(T)$ and $\rho_s(T)$ for the
parameters corresponding to the Melinte {\it et al} M242
sample\cite{melinte}. As can be seen, $\rho_s$ in particular has a
dramatic $T$ dependence, and essentially vanishes for $T>6K$. This
corroborates the earlier observation that the spin-reversed gap
collapses to $E_Z$ at around the same temperature, as seen in Figure 4.

Having extracted the parameters $M_0(T)$ and $\rho_s(T)$, it is now
easy to calculate the effects of spin fluctuations in the leading
large-$N$ approximation from equations
(\ref{scales},\ref{scaleo}). Figure 1 (in the Introduction) presents
the results for the Khandelwal {\it et al} data\cite{barrett}. As can
be seen, the agreement between the theory and experiment is now quite
good, indicating that the proper treatment of long-wavelength thermal
spin fluctuations has remedied most of the defects of the CFHF
prediction. Turning to the M242 sample of Melinte {\it et
al}\cite{melinte}, a similar dramatic improvement in the agreement
between theory and experiment is seen in Figure 2 (in the
Introduction). There are still some discrepancies in both the figures,
but they never amount to more than $10-15\%$.

One might wonder how important it is to keep the temperature
dependence of the parameters $M_0$ and $\rho_s$. Figure 8 shows the
prediction of the large-$N$ approach using the values of $M_0$ and
$\rho_s$ computed at $T=0$ for the parameters corresponding to the
Khandelwal {\it et al} data\cite{barrett}. This fit ignores
finite-temperature single particle excitations. It is clear that while
this prediction is somewhat better the the CFHF result for lower
temperatures, it is {\it worse} beyond $6 K$. Also, this prediction is
uniformly worse than the one including both CFHF and spin fluctuation
effects. 
\begin{figure}
\narrowtext
\epsfxsize=2.4in\epsfysize=2.4in
\hskip 0.3in\epsfbox{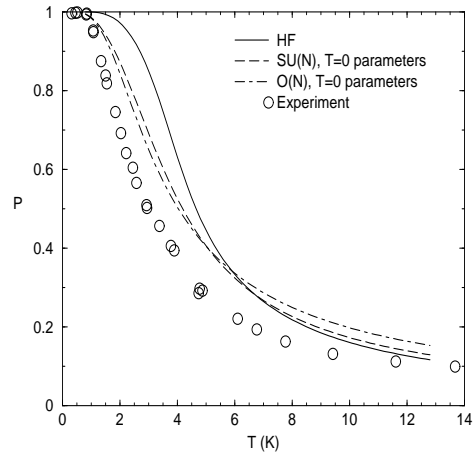}
\vskip 0.15in
\caption{The theoretical prediction for the Khandelwal {\it et al} 
data\cite{barrett} based on a zero temperature fit to the parameters
of the CQFM. The thickness parameter is assumed to be
$\lambda=1.5l$. Circles represent the experimental data\cite{barrett}.
\label{fig8}}
\end{figure}

Clearly, the reason is that this fit does not incorporate the
finite-temperature fermionic single-particle effects that renormalize
the spin stiffness substantially downwards beyond about $3K$, as seen
in Figure \ref{fig7}.  It must be concluded that the temperature dependence of
the parameters $M_0$ and $\rho_s$ is crucial, and that {\it both
single-particle and thermal spin wave effects must be treated
correctly} if one is to accurately predict the spin polarization.

\section{Conclusions, Caveats,  and Open Questions}

In summary, an approximate analytical method for computing the
temperature dependence of the polarization for an arbitrary fractional
quantum Hall state has been presented, and illustrated for $\nu=1/3$
and $2/5$.  It consists of two steps. The first step is a
finite-temperature Composite Fermion Hartree-Fock approximation on the
CF Hamiltonian. This already incorporates important effects on the
polarization resulting from particle-hole excitations, and is
sufficient to produce all the correct qualitative features, such as
the nonmonotonicity of the polarization for $2/5$\cite{exact}. For the
spontaneously polarized $1/3$ state, however, there is a substantial
discrepancy between this prediction and experiment at intermediate
temperatures due to the incorrect treatment of spin fluctuations. The
second step, for the $\nu=1/3$ state, is to map the long-wavelength
low-energy physics on to the continuum quantum
ferromagnet\cite{largen}.  The parameters that enter the effective
theory are extracted from the CFHF approximation.  Finally, the
approximate solution to continuum quantum ferromagnet in the large-$N$
approach\cite{largen} is used to produce a prediction for the spin
polarization which incorporates the effects of both single particle
and spin wave excitations. Note that the reduction of the spin
polarization due to single-particle and spin wave effects is not
additive in the two effects: the parameters $M_0$ and $\rho_s$
extracted from the CFHF approximation encode information about
single-particle excitations in a very nonlinear way. The resulting
prediction is in quite good agreement with experimental data, as seen
in Figures 1 and 2.  The comparison between theory and experiment was
made using the model potential of equation (\ref{zdspot}) for
illustrative purposes, with a thickness parameter $\lambda=1.5l$,
which ought to be in the physical regime for most
samples\cite{morf}. However, the above two-step procedure of first
carrying out the CFHF, and then mapping the problem on to the CQFM,
can be executed for {\it any} potential $v(q)$.

A number of caveats must be noted at this point. In order for the
mapping to the CQFM to make sense, spin wave excitations must be the
only low energy spin-flip excitations. However, as $T$ increases,
other spin-flip excitations also become relevant (such as a transition
from the $\ua$-spin $n=0$ CF-LL to the $\da$-spin $n=1$
CF-LL). Another complicating issue is that while the CQFM assumes the
spin wave dispersion to be quadratic at all $q$, in reality it turns
over and asymptotes at the spin-reversed gap $\Delta_{SR}$. Both these
considerations suggest that the mapping is to be trusted only in the
regime where $T\le\Delta_{SR}(T)$. One can verify from Figure 5 that
this corresponds to $T\le 6K$ or so, which fortunately includes the
most interesting region of the data. It is somewhat puzzling that the
prediction from the mapping works well to twice this
temperature. Finally, $1/N$ corrections can be expected to the leading
large-$N$ prediction, which reduce the magnetization slightly beyond
the leading large-$N$ result (see, for example, Timm {\it et
al}\cite{largen}). These corrections would improve the agreement
between the theory and the data at low $T$.

An important open problem is the development of a formalism in which
one can systematically integrate out fermions, leaving behind a theory
of the low-energy excitations, perhaps along the lines of
Ref. [15]. Such excitations need not be restricted to spin wave
excitations, but may perhaps include other spin-flip modes or even
density modes, depending on the temperature and the system. Such a
theory is necessary to address the effects of thermal spin
fluctuations on the temperature dependent spin polarization of
fractional Hall states which are {\it not} fully polarized in the
absence of a Zeeman coupling. The spin-flip modes of such systems do
not have the simple quadratic CQFM form even at small $q$ (for
example, the spin-flip collective mode of the fully polarized $2/5$
state starts with a {\it negative} quadratic term\cite{me-us}). A
further interesting application would be to the interplay of spin and
density fluctuations in the compressible fractional Hall
states\cite{hlr,shankar12}.

\section{Acknowledgements}

It is a pleasure to thank R. Shankar, J. K. Jain, H. A. Fertig,
S. M. Girvin, C. Timm, S. E. Barrett, and N. Freytag for helpful
conversations, and the latter two for sharing their raw data. The
author also wishes to thank the National Science Foundation for
partial support (under grant DMR 0071611), and the Aspen Center for
Physics for a stimulating workshop during which some of this work came
to fruition.


\begin{thebibliography}{100}
\bibitem{itinerant} Note that neither of the above formulas agrees
with the result for itinerant noninteracting electrons at $\nu=1$,
namely, $\Phi_M=\tanh(h/4)$. However, since we will be using the
large-$N$ forms with nonzero $r$ later, we have to use the above forms
for consistency.
\end{thebibliography}
\end{document}